\documentclass[12pt]{iopart}
\usepackage{graphicx}
\usepackage{iopams}
\usepackage{epsfig}

\begin{document}
\title{Modelling radiation emission in the transition from the classical to the quantum regime}
\author{J L Martins$^1$, M Vranic$^1$, T Grismayer$^1$, J Vieira$^1$, R A Fonseca$^{1,2}$, L O Silva$^1$}

\address{$^1$ GoLP/Instituto de Plasmas e Fus\~ao Nuclear, Instituto Superior T\'ecnico, Universidade de Lisboa, 1049-001 Lisbon, Portugal}
\address{$^2$ DCTI/ISCTE Instituto Universit\'{a}rio de Lisboa, 1649-026 Lisboa, Portugal}
\ead{jlmartins@ist.utl.pt}
\begin{abstract}
An emissivity formula is derived using the generalised Fermi-Weizacker-Williams method of virtual photons which accounts for the recoil the charged particle experiences as it emits radiation. It is found that through this derivation the formula obtained by Sokolov \emph{et al} using QED perturbation theory is recovered. The corrected emissivity formula is applied to nonlinear Thomson scattering scenarios in the transition from the classical to the quantum regime, for small values of the nonlinear quantum parameter $\chi$. Good agreement is found between this method and a QED probabilistic approach for scenarios where both are valid. In addition, signatures of the quantum corrections are identified and explored.
\end{abstract}
\submitto{PPCF}
\section{Introduction}

Increasing available laser power and high-energy electron beams (reaching few GeV; for e.g. \cite{lwfa1p45GeV,lwfa2GeV,lwfa4GeV}) obtained with plasma-based accelerators allow us to reach new regimes in laser-beam scattering in all-optical configurations \cite{prl_marija,malka_compton,lwfa_compton}.  At sufficiently high laser intensities, radiation damping becomes non-negligible and needs to be taken into account in the electron dynamics through the radiation reaction force. For even greater intensities or for extremely relativistic electrons and moderate intensities, the energy loss in the emission of a single photon can be so big that the trajectory is no longer continuous and radiation reaction needs to be described with quantum electrodynamics. The transition to the quantum regime can be parameterised by:
\begin{eqnarray}
\chi = \frac{|F_{\mu\nu} \, p^{\nu}|}{E_{crit} mc},
\label{eq:chi}
\end{eqnarray}
where $E_{crit} = m^2 c^3 / (e \hbar)$ is the Schwinger critical field, $F_{\mu\nu}$ is the electromagnetic wave field-strength tensor and $p^{\nu}$ is the momentum four-vector. This parameter is approximately equal to the ratio between the field amplitude in the rest frame of the electron and the Schwinger critical field for very relativistic electrons.
When $\chi \lesssim 1$ emission occurs through Compton scattering and, as the emitted energy becomes comparable to the electron energy, the trajectory gets stochastic; at $\chi ~\sim 1$ processes such as pair production become significant. However, for $\chi \ll 1$, the trajectory of the electron can still be described classically if one accounts for the radiation reaction force. The emission process is Thomson scattering, the classical analog of Compton scattering. In this paper, the definition for nonlinear Thomson/Compton scattering from ref. \cite{DiPiazza_review} is used.

Another important parameter in describing the nature of electron-laser scattering is the normalised vector potential of the laser $a_0$. For very intense lasers, $a_0 \gg 1$ the electron dynamics in the laser is nonlinear and radiation is emitted at multiple harmonics (nonlinear Thomson \cite{nonlinThomson1,nonlinThomson2}/Compton \cite{nonlinCompton1,nonlinCompton2} scattering). In this work, the regime of radiation emission with $\chi \ll 1$ to $< 1$ and $a_0 \gg 1$ and $\gamma \gg 1$ will be studied, where $\gamma$ is the Lorentz factor of the electron. In it, radiation damping is non-negligible but can be described classically and the higher harmonics of nonlinear Thomson scattering extend to energies approaching the energy of the electron, such that the classical emissivity formula derived from the Li\'enard-Wiechert potentials may not be valid anymore. The question then arises of how to model radiation emission in this scenario.

Several techniques have been used, such as semiclassical calculations \cite{semiclass_1,semiclass_2,semiclass_3}, QED perturbation theory \cite{sokolov}, the introduction of functions that correct the equation of motion and the radiated power/spectrum \cite{gfactor_1,gfactor_2,sokolov} and Monte-Carlo methods, either based on the cross sections for Compton scattering \cite{linCompton_1,linCompton_2} or on the emission probability function and spectrum of synchrotron radiation combined \cite{qed_pic_1,qed_pic_2,qed_pic_3,qed_pic_4,qed_pic_5}.

In this paper we show how to derive a quantum corrected emissivity formula for arbitrary observation directions using the generalised method of Fermi-Weizacker-Williams, first developed by Lieu \& Axford \cite{lieu}. This formula is implemented in the post-processing radiation diagnostic code JRad \cite{spie_jrad}. It is then used to explore the changes that occur in the Thomson scattering spectrum in the transition from the classical to the quantum regime. 

This paper is structured as follows: in section I, a quantum corrected emissivity formula is derived. In section II, comparisons are shown between the spectrum computed with this technique and the spectra obtained through a QED probabilistic approach \cite{qed_pic_4} for synchrotron radiation. In section III, we explore nonlinear Thomson scattering in the transition from the classical to the quantum regime with JRad and trajectories obtained from the integration of the equation of motion with the radiation reaction force \cite{RR_PIC_marija}. We show that under certain conditions a unique signature of the quantum corrections is observed, which is not seen if only the radiation damping is taken into account in the trajectory. In section IV we state the main conclusions of the paper.

\section{Radiation emissivity with quantum corrections}
\label{}

The method of Lieu \& Axford consists of an extension of the method of virtual photons (FWW method) \cite{lieu}. This method was originally applied to problems such as bremsstrahlung radiation, where the electron was considered to be in uniform motion. The application of the FWW method to scenarios such as an electron gyrating in a uniform and static magnetic field is not possible, since there is no single rest frame for the electron. Lieu $\&$ Axford have proposed a method to overcome this issue \cite{lieu}. Their solution relies on splitting the electron trajectory into a series of infinitesimal segments in which the electron velocity is approximately constant and therefore the method of virtual photons can be applied. They then derive the classical results of synchrotron radiation by determining the Thomson scattering spectrum in this series of instantaneous rest frames, transforming the emissivities back to the laboratory and then adding them coherently. Quantum corrections associated with the recoil of the electron during the emission process are added by replacing the Thomson scattering  cross section by the Compton cross section and $\omega$ by $\omega/\eta$, where $\eta$ is the Compton shift. Though Lieu \& Axford have extended their work to three-dimensional scenarios \cite{lieu3D}, their 3D emissivity formula only reduces (in the absence of quantum corrections) to the classical result for certain angles of observation.

In the following, it is shown that by introducing an additional generalisation to the method of Lieu \& Axford, it is possible to obtain a quantum corrected emissivity formula which reduces to the three-dimensional classical emissivity in the limit of negligible Compton shift for arbitrary directions of observation. For the sake of completeness, the essential steps in the derivation are presented here; for further details, the reader is invited to check ref. \cite{lieu,joanalm_thesis}.

Firstly, a coordinate transformation is performed in the laboratory frame such that the segment to be analyzed has its beginning at the origin. A Lorentz transformation is then performed to the instantaneous rest frame of the electron for this segment. In this frame, the component of the Poynting flux for the frequency $\omega'$, $S'(\omega ')$ is given by:
\begin{eqnarray}
S'(\omega ') = c \left| \frac{1}{2\pi} \int \mathbf{E'} e^{i\omega't'} dt'\right|^2,
\label{eq:Sw}
\end{eqnarray}
where $c$ is the velocity of light, $\mathbf{E'}$ is the incident electric field, $t'$ is time and the primed variables refer to quantities in the instantaneous rest frame.

In this infinitesimally small segment the electron will remain non-relativistic and the electric field can be related to the electron velocity from the equation of motion $m \dot{\mathbf{v}}_{\perp}' = e \mathbf{E}'$, where $m$ and $e$ are the mass and charge of the electron, and $\dot{\mathbf{v}}_{\perp}'$ is the acceleration of the electron. Assuming that the field can be decomposed into a series of plane waves, the Poynting flux for the segment then becomes after integration in time: 
\begin{eqnarray}
\Delta S'(\omega ') = c  \left( \frac{\omega' m}{2\pi e} \right)^2 \left| \mathbf{v_{\perp}'}(t') e^{i\omega't'}\Delta t' \right|^2,
\label{eq:deltaSw}
\end{eqnarray}
where $\Delta t'$ is the time the electron takes to cross the infinitesimal segment.

The emissivity at a frequency $\omega '$ (energy radiated at a frequency $\omega '$ per unit of frequency and per unit of solid angle) in the infinitesimal segment of trajectory, $\Delta \alpha '$, is given by the product of the incident radiation flux and the Compton cross section $(d\sigma'/d\mathbf{\Omega}')_C = [e^2 / (m c^2)]^2 \,\eta' \left| \bepsilon_{out}' \cdot \bepsilon_{in}' \right|^2$, where $\bepsilon_{in}'$ and $\bepsilon_{out}'$ are the polarisation vectors of the incident and outgoing photons. To account for the Compton shift corrections, $\omega'$ is replaced by to $\omega'/\eta'$ and the associated substitution:
\begin{eqnarray}
\alpha = \frac{d^2 I}{d\omega d\Omega} \longrightarrow \alpha\, \eta^2 = \frac{d^2 I}{d\omega d\Omega} \eta^2
\end{eqnarray}
is made.
The emissivity in the infinitesimal rest frame with the quantum correction is then given by:
\begin{eqnarray}
\Delta \alpha'(\omega ',\mathbf{\Omega}') = \frac{e^2 {\omega'}^2}{4 \pi^2 c^3} \left| \frac{(\boldsymbol{\epsilon_{out}'} \cdot \boldsymbol{\epsilon_{in}'})}{\sqrt{\eta'}}\,  \mathbf{v_{\perp}'}(t') e^{i\frac{\omega'}{\eta'} t'} \Delta t' \right|^2.
\label{eq:deltaAlphaPrime2}
\end{eqnarray}
To transform this result back to the laboratory frame it is useful to note that the dot product $\boldsymbol{\epsilon_{out}'} \cdot \boldsymbol{\epsilon_{in}'}$, $\mathbf{v_{\perp}'}(t') \Delta t'$ and (for very relativistic particles) $\eta'$ are invariants, as pointed out by Lieu \& Axford \cite{lieu}. To facilitate the calculations, these two quantities are determined in the rest frame of the particle but expressed in terms of the variables in the lab frame \cite{lieu}. The transformation of the time and frequency yields $i\omega't' = i\omega (t - \mathbf{n}\cdot\mathbf{v}t/c)$ and when transforming back from the simplified reference frame to the original laboratory frame the term $\mathbf{n}\cdot\mathbf{v}t$ in the exponent yields $\mathbf{n}\cdot\mathbf{r}$.

To obtain the emissivity in the laboratory, the contributions to the total radiation spectrum from each segment $\Delta \alpha$ need to be Lorentz transformed to this frame and added coherently, by summing them before taking the square of the modulus \cite{lieu} in equation (\ref{eq:deltaAlphaPrime2}):
\begin{eqnarray}
\alpha = \frac{d^2 I}{d\omega d\Omega} = \frac{e^2 \omega^2}{4 \pi^2 c^3} \left| \int \frac{\boldsymbol{\epsilon_{out}} \cdot \mathbf{v_{\perp}}}{\sqrt{\eta}} e^{i\frac{\omega}{\eta} (t-\mathbf{n}\cdot\mathbf{r}/c)} d t \right|^2.
\label{eq:alpha1}
\end{eqnarray}

To facilitate the calculations, the direction of the velocity can be taken to be in the positive $x$ direction and spherical coordinates will be used. A natural choice of unity vector in the observation direction for the outgoing radiation is then $\boldsymbol{n}' = \mathbf{e_r'}/|\mathbf{e_r'}|$, i.e. $\mathbf{n}' = (\cos\theta ',\sin \theta '\sin \phi ',\sin\theta '\cos \phi ')$, where $\theta'$ is the angle between the wave vector of the outgoing radiation and the velocity and $\phi'$ is the angle between the two transverse components of the observation direction, z and y. 
The polarisation vectors can then be given by the other unity vectors in the spherical coordinates triad, $\mathbf{e_{\theta}'}$ and $\mathbf{e_{\phi}'}$, which in terms of the laboratory frame angles gives:
\begin{eqnarray}
\boldsymbol{{\epsilon_{out}'}^1} &=& \left( \frac{-\sin \theta}{\gamma (1-\beta \cos \theta)}, \frac{(\cos \theta - \beta) \sin \phi}{(1 - \beta \cos \theta)}, \frac{(\cos \theta - \beta) \cos \phi}{(1 - \beta\cos\theta)} \right) \\
\boldsymbol{{\epsilon_{out}'}^2} &=& (0,\cos \phi,-\sin \phi).
\end{eqnarray}

The displacement in the rest frame can be expressed in terms of the laboratory quantities:
\begin{eqnarray}
(\Delta x',\Delta y' ,\Delta z') &=& (\gamma (v_x - v)\Delta t,v_y \Delta t,v_z \Delta t) \nonumber,
\label{eq:displac_3D}
\end{eqnarray}
where $\gamma$ is the Lorentz factor of the electron, $v_x$, $v_y$ and $v_z$ are the components of the velocity of the electron in the laboratory, in the trajectory infinitesimal segment. 

Substituting the rest frame displacement and the outgoing radiation polarisation vectors into equation (\ref{eq:deltaAlphaPrime2}) the specific emissivities in the polarisation directions $\boldsymbol{{\epsilon_{out}'}^1}$ and $\boldsymbol{{\epsilon_{out}'}^2}$, denoted by $\alpha_1$ and $\alpha_2$ respectively, can be computed. The two contributions can be summed inside the modulus since they are perpendicular to each other and the complete emissivity formula is obtained:
\begin{eqnarray}
\frac{d^2 I}{d\omega d\Omega} = \frac{e^2 \omega^2}{4 \pi^2 c^3} \left|\, \int \frac{\mathbf{n}\times (\mathbf{n}\times\boldsymbol{\beta})}{\sqrt{\eta}} \exp\left[i\frac{\omega}{\eta} (t - \mathbf{n}\cdot\mathbf{r}/c)\right] d t \right|^2
\label{eq:emissivity_QC}
\end{eqnarray}
This formula (which recovers the result of Sokolov \emph{et al} \cite{sokolov}) reduces to the classical emissivity in the limit $\eta \rightarrow 1$, whereas the three-dimensional result of Lieu \& Axford \cite{lieu3D} did not when $\phi \ne 0$. The emissivity obtained here is then applicable to an arbitrary observation direction. The changes in the calculations above which allowed this generalisation were two: setting the unit vector of observation direction $\mathbf{n'}$ to $\boldsymbol{n}' = \mathbf{e_r'}/|\mathbf{e_r'}|$ and extending the perpendicular displacement vector $\Delta \mathbf{r'} = \mathbf{v'}_{\perp} \Delta t'$ to three dimensions. We also note that in doing the integration by parts in the calculation of $\alpha_1$, the term
\begin{eqnarray}
\left[ \frac{\exp (i\frac{\omega}{\eta} t)}{i\frac{\omega}{\eta}}\exp\left(-i\frac{\omega}{\eta} \mathbf{n}\cdot\mathbf{r}/c\right)\right]_{-\infty}^{+\infty}
\end{eqnarray}
was neglected since the modulus of a complex exponential is $\le 1$ and it is assumed that $\omega \gg 1$. 

Equation (\ref{eq:emissivity_QC}) can be implemented in post-processing radiation diagnostic codes such as JRad \cite{spie_jrad}, which was the one used in this work. The particle trajectories can be obtained from particle-in-cell codes or from the integration of the equation of motion, including the radiation reaction force.   

\section{Comparison with QED results}

As a test to the quantum corrected diagnostic (JRad-QC), the synchrotron radiation of an ultra-relativistic electron in an ultra-intense magnetic field was computed with both JRad-QC and with OSIRIS-QED \cite{osiris,qed_pic_4}. The latter calculates the QED probability of radiation emission and determines the photon energy to be emitted using a Monte-Carlo method and the QED synchrotron radiation spectrum.

In this benchmark the spectrum of an electron with $\gamma = 200$ subject to a magnetic field of $B = 5.7 \times 10^{10}\,\mathrm{G}$ (which corresponds to $\chi = 0.26$) is compared with the result from OSIRIS-QED without damping. The case without damping is presented for reference since the theoretical radiation spectrum is known, and therefore can easily be used for benchmarking purposes. In this case, the damping is turned off in OSIRIS-QED, hence the spectrum comparison tests solely whether the quantum correction in the model of Lieu \& Axford reproduces the results obtained for photon production according to a QED calculation of nonlinear Compton scattering. 

\begin{figure} [h!]
\begin{center}
\includegraphics[height=4.5cm]{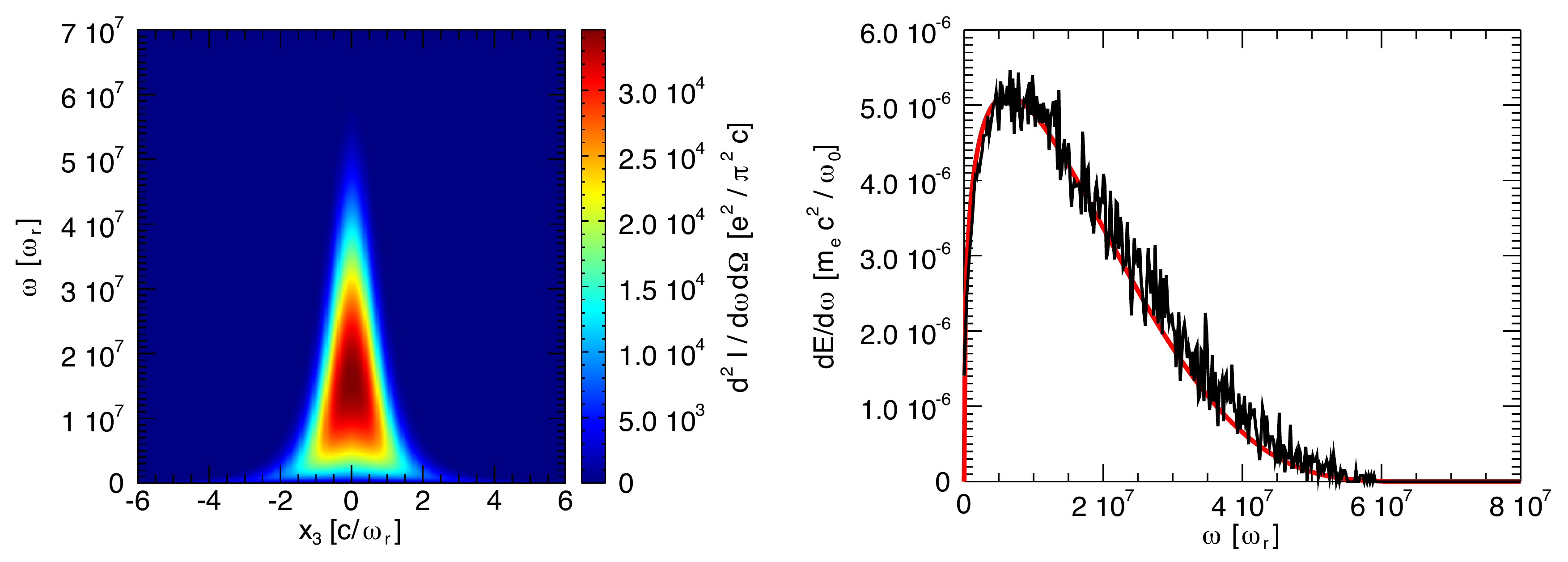}
\end{center}
\caption{\label{fig:synch_bench} (left) Spectrum determined over a line perpendicular to the synchrotron trajectory plane determined from the JRad diagnostic with quantum corrections (JRad-QC). (right) The red line represents the synchrotron spectrum obtained with JRad-QC and the black line is the result obtained with OSIRIS-QED.}
\end{figure} 

For this benchmark, an analytical trajectory was produced according to the classical equations of motion for an electron in a static magnetic field (see for e.g. \cite{Rybicki}). The post-processing diagnostic with quantum corrections was used to obtain the emitted spectrum as a function of solid angle along a line perpendicular to the trajectory plane, at a distance of about $150 \,c/\omega_r$ from the gyration circumference (the precise value is not relevant as long as it is in the far field), where $\omega_r = 1.88 \times 10^{15}$ rad/s is the normalisation frequency, equivalent to $E = \hbar \omega_r \simeq 1.24 \,\mathrm{eV}$. This spectra was then integrated in solid angle to obtain $dE/d\omega$ (seen in red in Figure \ref{fig:synch_bench}). The result shows excellent agreement between the OSIRIS-QED and the JRAD-FWW spectra (which had been shown to be in agreement with the theoretical result for synchrotron in the QED regime). We observe that comparisons with QED calculations need to be done with careful consideration since the method is only valid for ultra relativistic particles and sufficiently high fields \cite{qed_synch_rad_1,qed_synch_rad_2,qed_synch_rad_3}.

\section{Transition from the classical to the quantum regime}
 
 To explore the nonlinear Thomson scattering spectrum in the transition to the quantum regime, trajectories from ultra-relativistic electrons colliding with plane waves and laser pulses were produced by numerical integration of the equation of motion with a Runge-Kutta of $4^{th}$ order \cite{RR_PIC_marija}. The radiation diagnostic JRad-QC / JRad was then used to compute the radiation spectra at line detectors in the far-field region with/without quantum corrections.
  
 To investigate when quantum effects start to affect significantly the radiation spectrum a series of spectra has been computed for the case of a circularly polarised plane wave with normalised vector potential of $a_0 = 0.88, \,\, 1.77, \,\, 7.07$ scattering off an electron with $\gamma=10^3$ (travelling in the $+ x_1$ direction) in counter-propagating geometry. The plane wave is preceded by a rising ramp but the electron energy is ajusted so that $\gamma \simeq 10^3$ at the beginning of the flat part of the wave, which lasts for $t_{flat} \simeq 38 \,\mathrm{T_0} \simeq 127 \,\mathrm{fs}$, where $T_0$ is the wave period.
The virtual detector is a line at $x_3 = 0$ and the normalisation frequency is the same as the frequency of the monochromatic wave $\omega_0 = 1.88\times 10^{15}$ rad/s (which corresponds to $1 \,\mu\mathrm{m}$ wavelength or $E = \hbar \omega_0 \simeq 1.24 \,\mathrm{eV}$). 

Two effects are readily observed in the spectra in Figure \ref{fig:cp_plane_a0_0p88_1p77_7p07}. Firstly, in the case of $a_0 = 0.88$ there is a shift in the frequency of the harmonics. Secondly, in all the cases, a reduction of the radiated energy at higher frequencies is observed, as expected \cite{semiclass_3,semiclass_2}.
\begin{figure} [h!]
\begin{center}
\begin{tabular}{c}
\includegraphics[height=8.0cm]{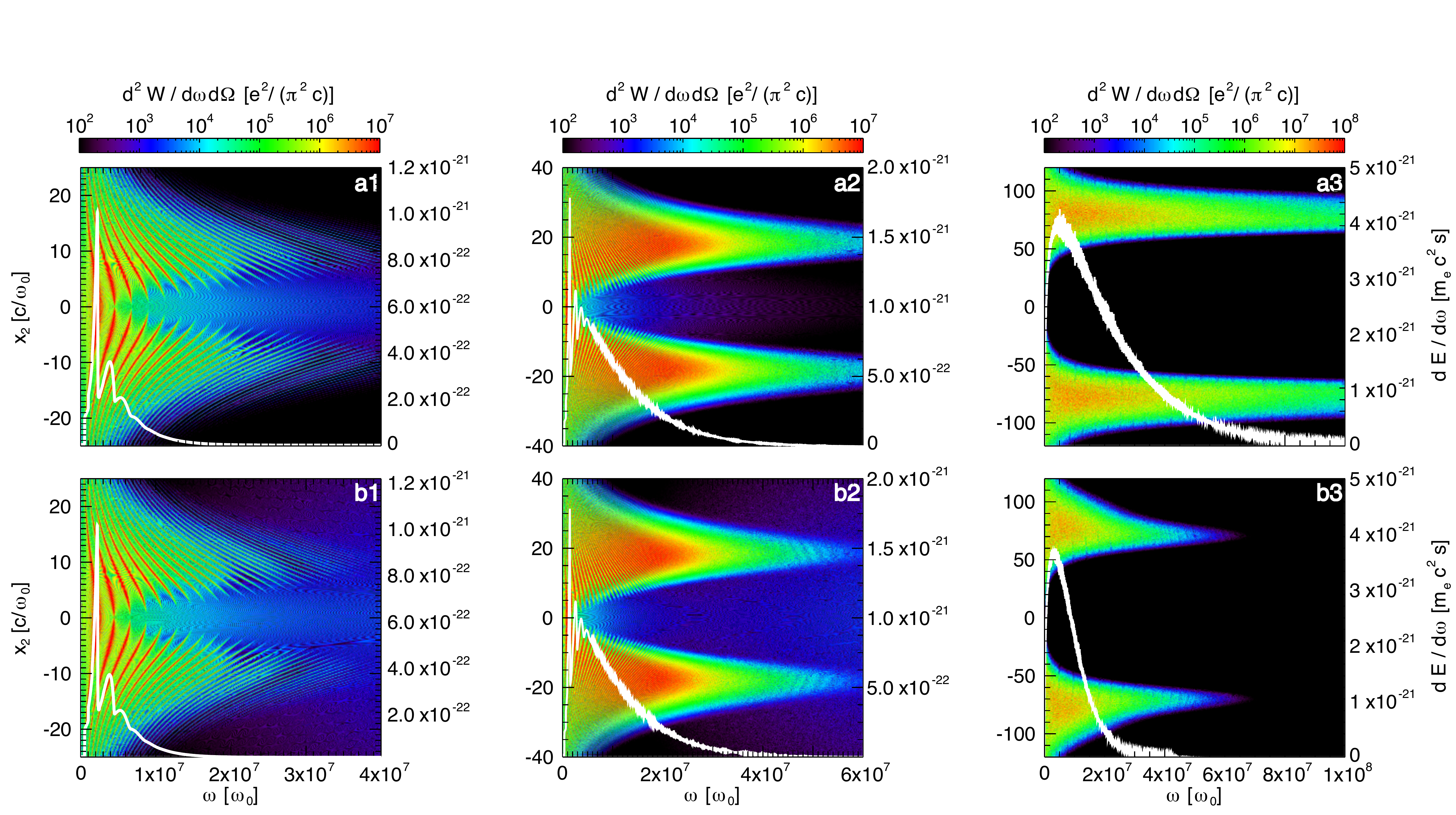}
\end{tabular}
\end{center}
\caption{ \label{fig:cp_plane_a0_0p88_1p77_7p07} Spectra of the scattering of a circularly polarised plane wave with $a_0 = 0.88$ (1), $a_0 = 1.77$ (2) and $a_0 = 7.07$ (3). Plots labeled with (b) refer to the calculation with quantum correction (JRad-QC) and those labeled with (a) to the classical emissivity calculation (JRad). The white lines represent the spectra integrated in solid angle assuming cylindrical symmetry.}
\end{figure} 
The energy lost by the electron over the interaction time for the case of $a_0 = 7.07$ was about $21 \%$, which gives an average energy loss rate of $0.55 \%$ per period of oscillation. This slow energy loss rate is the reason behind the almost symmetric profile in the spectra, which would not be expected if the electron would change its energy by a significant amount during an oscillation period. It is then possible to estimate the total energy captured by the detector by doing an integration in solid angle and in photon energy, assuming cylindrical symmetry around the direction defined by the initial momentum direction of the electron, in this example the line in the $+x_1$ direction, at $x_2 = 0$ and $x_3 = 0$. Doing so, one obtains 271 keV from the integration of the detector without quantum corrections, 104 keV for the detector with quantum corrections, and 109 keV from the direct measure of the energy loss by the particle in its trajectory. The energy captured in the detector with the quantum corrections is very close to the value measured in the track, which further supports the validity of the formula for the quantum corrected emissivity. 

Regarding the other values of $a_0$ investigated, the energy captured in the detector with quantum corrections is also closer to the energy difference measured in the track but the difference is smaller, which could be anticipated from the fact that the observed differences in the spectrum are also less significant than in the highest $a_0$ case.

\section{Laser pulse scattering}

The scattering of a linearly polarised laser pulse from an electron with $\gamma=1030$ (travelling in the $+ x_1$ direction) in counter-propagating geometry was also explored, in a configuration similar to ref. \cite{prl_marija}. In the scenario studied here, the laser pulse has a $1 \,\mu\mathrm{m}$ wavelength, 26.5 fs duration and peak normalised vector potential of $a_0 = 10, \,\, 20, \,\,30$ and is linearly polarised in the $x_2$ direction.
\begin{figure} [h!]
\begin{center}
\begin{tabular}{c}
\includegraphics[height=8.0cm]{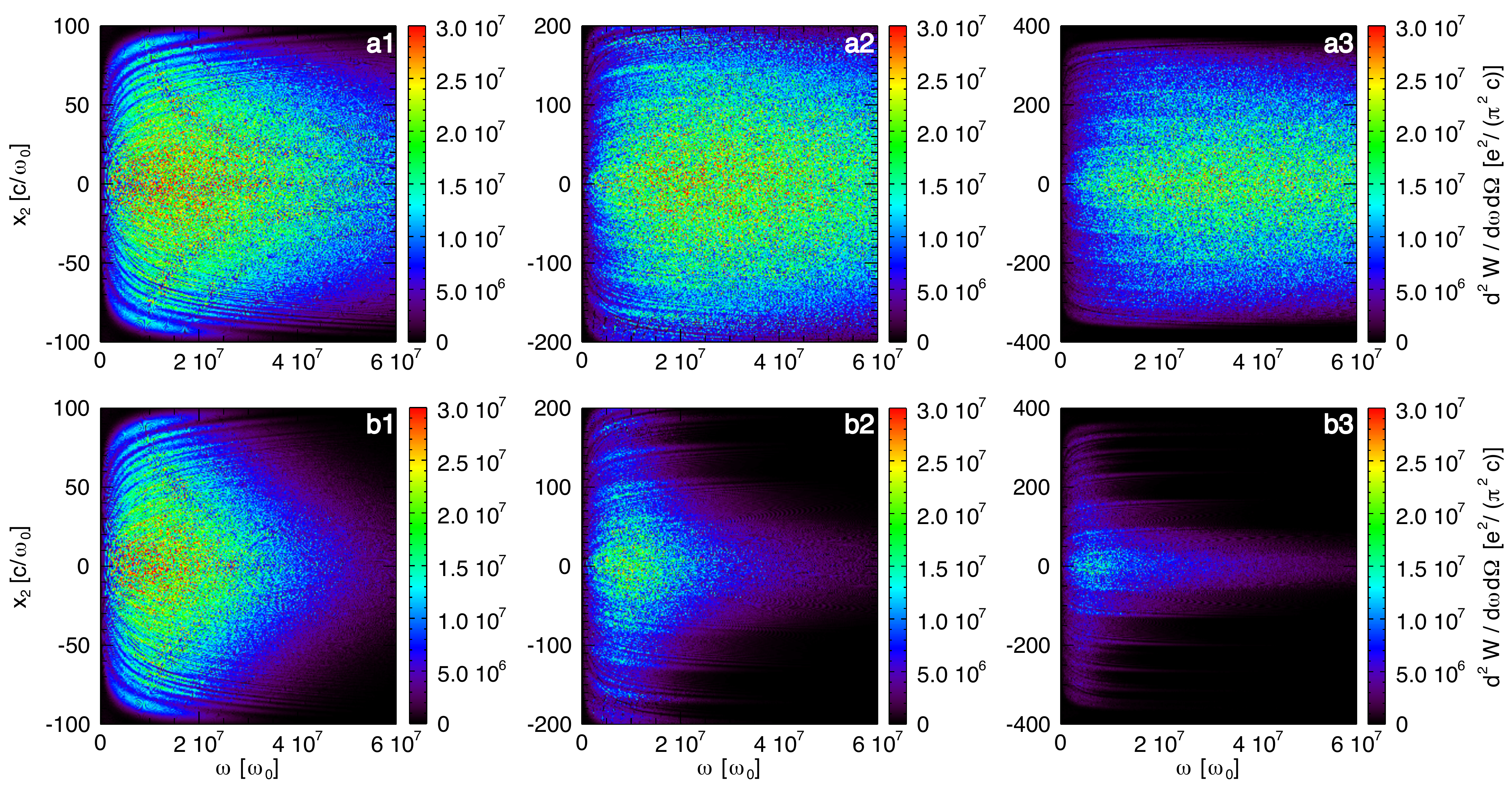}
\end{tabular}
\end{center}
\caption{ \label{fig:cp_plane_a0_0p88_1p77_7p07} Spectra of the scattering of a linearly polarised laser pulse with $a_0 = 10$ (1), $a_0 = 20$ (2) and $a_0 = 30$ (3). Plots labeled with (b) refer to the calculation with quantum correction (JRad-QC) and those labeled with (a) to the classical emissivity calculation (JRad).}
\end{figure} 
\begin{figure} [h!]
\begin{center}
\begin{tabular}{c}
\includegraphics[height=9.0cm]{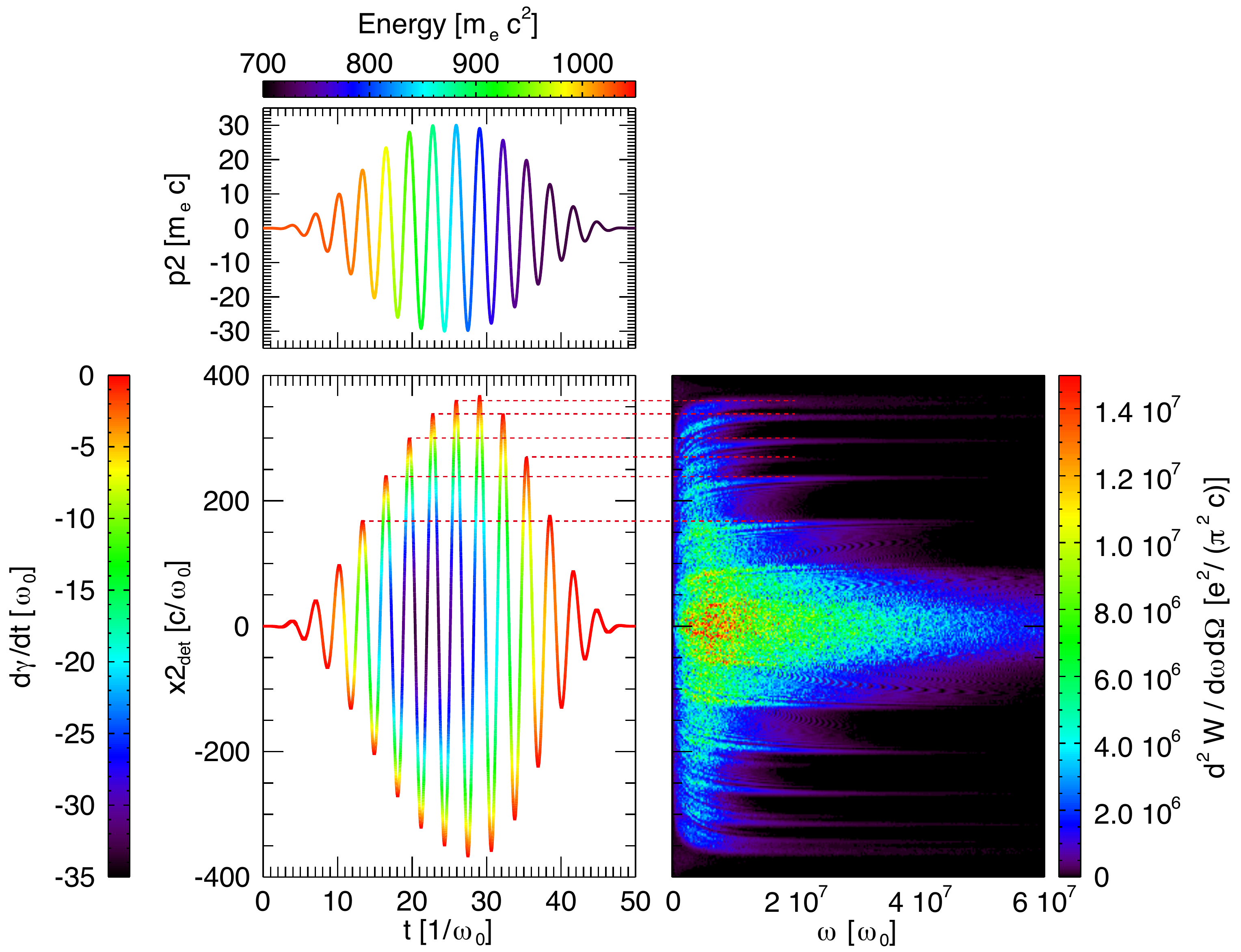}
\end{tabular}
\end{center}
\caption{ \label{fig:spec_pulse_a0_30_and_track} Evolution of the electron transverse momentum ($p2$) (left top pane) while interacting with a laser pulse with $a_0 = 30$ and of the position in the detector to which it points to (bottom left pane). Line spectrum from the scattering of the laser pulse off the electron (right pane). The dashed lines show the peaks in the spectrum originate from points in the trajectory where $d\gamma/dt = 0$, which correspond to peaks in the transverse momentum.}
\end{figure} 
The spectra computed for a line positioned at $x_1 = 10^4 \,c/\omega_0$ and $x_3 = 0 \,c/\omega_0$ with and without quantum corrections are depicted in Figure \ref{fig:cp_plane_a0_0p88_1p77_7p07}. From the results, it can be seen that as the peak $a_0$ increases the shape of the spectrum with quantum corrections changes more and more significantly, specially at higher angles of observation (higher $x_2$ values in the detector axis). In addition, at the highest $a_0$, spike structures are clearly seen. These are correlated with the maxima in the transverse momenta, where the derivative of $\gamma$ is zero (see Figure \ref{fig:spec_pulse_a0_30_and_track}), and exist also in the classical case. However, the quantum corrections lead to a stronger decrease in radiated energy between the spikes, making them appear more clearly.

The spectrum changes are more complex and seem more dependent in angle in the laser pulse scattering setup compared with the plane wave scenario. This will be explored in more detail in a future publication. 

\section{Conclusions}

In this paper it was shown that by extending the generalised FWW method of Lieu \& Axford, the emissivity with quantum corrections due to the electron recoil can be derived, which had been obtained by a different approach by Sokolov \emph{et al} \cite{sokolov}, through QED perturbation calculations.

The quantum corrected emissivity was implemented in the numerical diagnostic JRad \cite{spie_jrad} and used to determine the nonlinear scattering spectrum of relativistic electrons with ultra-intense laser pulses and plane waves. It was found that only when the  quantum corrections are introduced does the energy captured in the virtual detector spectrum become consistent with the energy loss by the particle (as measured from the integration of its equation of motion).

The analysis of the scattering of ultra-intense linearly polarised laser pulses from relativistic electrons revealed that the changes in the transverse momenta and energy of the electron during the interaction led to a complex spectrum shape. For larger angles of observation, spike features are observed in the spatially resolved radiation spectrum, surrounded by regions of lower radiation emission. The shape changes reflect the changes in the Compton shift during the interaction, which were much more significant in the laser pulse scenario than in the plane wave case for the parameters used. This can be attributed to a combination of stronger radiation damping and bigger changes in the field amplitude experienced by the electron in the laser pulse scattering scenario with higher peak $a_0$ as compared to the plane wave cases.

\ack{This work was partially supported by the European Research Council (ERC − 2010 − AdG Grant 267841). The authors also wishes to acknowledge the computing facilities where the simulations/post-processing were done: the SuperMUC supercomputer (through PRACE), based in Leibniz research center, in Germany, and the cluster ACCELERATES in Instituto Superior T\'ecnico, in Lisbon, Portugal.}
	
\end{document}